\documentclass[aps,pra,twocolumn,groupedaddress,showpacs, showkeys]{revtex4}
\usepackage{amsmath}
\usepackage{amssymb}
\usepackage{graphicx}
\begin{document}
\title{Non-classical conditional probability and the quantum no-cloning theorem}
\author{Gerd Niestegge}
\affiliation{Fraunhofer ESK, Hansastr. 32, 80686 M\"unchen, Germany}
\email[] {gerd.niestegge@web.de}
\begin{abstract}
The quantum mechanical no-cloning theorem for pure states
is generalized and transfered to
the quantum logics with a conditional 
probability calculus in a 
rather abstract, though simple and basic fashion
without relying on a tensor product construction 
or finite dimension as required in other generalizations.
\end{abstract}
 

\maketitle

\noindent
\large
\textbf{1. Introduction}
\normalsize
\\[0,5 cm]
A pioneering result with far-reaching consequences in quantum information and communication theory is the no-cloning theorem, stating that unknown pure quantum states cannot be copied unless they are orthogonal \cite{wootters1982single, dieks1982communication, yuen1986amplification, werner2001quantum}. An interesting generalization is the no-broadcasting theorem for mixed states \cite{barnum1996noncommuting}.
Originally, both were proved in Hilbert space quantum mechanics, then extended to the C*-algebraic setting \cite{clifton2003characterizing} and later to finite-dimensional generic probabilistic models \cite{barnum2006cloning, barnum2007generalized} and to quantum logics \cite{:/content/aip/journal/jmp/50/10/10.1063/1.3245811}. In the latter case, only universal cloning is impossible, while the cloning of a small set or pair of states can be ruled out in the other cases. Though these results preclude the perfect cloning, the approximate or imperfect cloning of quantum states remains possible \cite{buvzek1996quantum, PhysRevA.57.2368, Kitajima2015}. 
In this paper, the (perfect) cloning of a small set or pair of states is considered in the setting of quantum logics with a conditional probability calculus \cite{niestegge2001non, niestegge2008approach}, including finite-dimensional as well as infinite-dimensional models.

A quantum logic $E$ is a purely algebraic structure for the quantum events (or propositions). It is quite common to use an orthomodular partially ordered set or lattice \cite{kalmbachorthomodular, beltrametti1984logic, beran1985orthomodular, ptak1991orthomodular}. 
States are then defined in the same way as the classical probability measures, and conditional probabilities are postulated to behave like the classical ones on compatible subsets of $E$. Note that a subset is called compatible if it is contained in another subset of $E$ forming a Boolean algebra (i.e., in a classical subsystem of $E$) \cite{brabec1979compatibility}. Some quantum logics entail unique conditional probabilities, many others don't. The classical Boolean algebras and the Hilbert space quantum logic (consisting of the closed subspaces or, equivalently, the selfadjoint projection operators) do and, in the latter case, conditionalization becomes identical with the state transition of the L\"uders - von Neumann (i.e., projective) quantum measurement process \cite{niestegge2001non}. Therefore, the quantum logics with unique conditional probabilities can be regarded as a generalized mathematical model of projective quantum measurement.
Examples that are neither classical Boolean algebras nor Hilbert space quantum logics (nor sublogics of them) can be constructed using the exceptional real Jordan algebras \cite{niestegge2001non, niestegge2008approach}. 

In this framework, a very special type of conditional probability emerges 
in the non-classical case \cite{niestegge2001non, niestegge2008approach}. 
It describes the probability for the transition from a past event $e$ to a future event $f$,
independently of any underlying state, and results from the algebraic structure 
of the quantum logic $E$. This probability exists only for certain event pairs $e$ and $f$. 
It exists for all events $f \in E$, if $e$ is a minimal event (\textit{atom}) in $E$. 
The states resulting in this way are called \textit{atomic}. They represent a generalization 
of the pure states in Hilbert space quantum mechanics.

After the early pioneering work by Birkhoff and von Neumann in 1936 \cite{birkhoff-vN36}, 
quantum logics have been studied extensively between 1960 and 1995 
\cite{piron1964axiomatique, varadarajan1968geometry, varadarajan1970geometry, Keller1980, soler1995characterization,
kalmbachorthomodular, beltrametti1984logic, beran1985orthomodular, ptak1991orthomodular}. 
Various forms of conditional probability have also been considered
\cite{gunson1967algebraic, bub1977neumann, friedman1978quantum, 
guz1981conditional, beltrametti1984logic, edwards1990conditional}.  
However, the quantum logics which possess unique conditional probabilities and particularly
the special type of the state-independent conditional probability 
have not received any attention before the author's work 
\cite{niestegge2001non, niestegge2008approach}.

Considering such a quantum logic, 
this special type of conditional probability 
is used in the present paper to prove, 
in a very basic fashion, the generalized no-cloning theorem for atomic states. 
A tensor product construction as used in the other approaches is not required. 
Instead, the embedding of two copies of $E$, which shall be 
compatible with each other, in a larger quantum logic $L$ is sufficient.

The paper is organized as follows.
The algebraic structure of the quantum logic is considered in section 2.
Section 3 then turns to states and briefly sketches 
the non-classical conditional probability calculus 
from Refs. \cite{niestegge2001non, niestegge2008approach}.
The main results are presented in sections 4 and 5.
The proof of the quantum mechanical no-cloning theorem  
rests upon the following two basic properties of the inner product in the Hilbert space: 
It is invariant under the unitary cloning transformation and
multiplicative for the Hilbert space tensor product. 
These properties are transferred
to the non-classical conditional probabilities in a certain way (Lemmas 1(a) and 2),
which then allows to mimic the quantum mechanical proof 
for the generalized no-cloning theorem (Theorem 1). 
In section 6, 
the ties to Hilbert space quantum mechanics are pointed out.
\newpage
\noindent
\large
\textbf{2. Compatibility in orthomodular}
\newline
\hspace*{0,5 cm}
\textbf{partially ordered sets}
\normalsize
\\[0,5 cm]
In quantum mechanics, the measurable quantities of a physical 
system are re\-presented by observables. Most simple are those 
observables where only the two discrete values 0 and 1 are 
possible as measurement outcome; these observables are called \textit{events} (or \textit{propositions})
and are elements of a mathematical structure called \textit{quantum logic}.

In this paper, a quantum logic shall be an 
orthomodular partially ordered set $E$ 
with the partial ordering $\leq$,
the orthocomplementation $'$, 
the smallest element $0$ and the largest element $\mathbb{I}$
\cite{kalmbachorthomodular, beltrametti1984logic, beran1985orthomodular, ptak1991orthomodular}.
This means that the following conditions are satisfied by all $e,f \in E$:
\\[0,2 cm]
%\hspace*{0,5 cm} 
(A) $ e \leq f$ implies $f' \leq e'$.
\\[0,2 cm]
%\hspace*{0,5 cm} 
(B) $(e')' = e$.
\\[0,2 cm]
%\hspace*{0,5 cm} 
(C) $e \leq f'$ implies $e \vee f$, the supremum of $e$ and $f$, exists.
\\[0,2 cm]
%\hspace*{0,5 cm} 
(D) $e \vee e' = \mathbb{I}$.
\\[0,2 cm]
%\hspace*{0,5 cm} 
(E) $f \leq e$ implies $e = f \vee (e \wedge f')$. \hspace*{0,5 cm}(orthomodular law)
\\[0,2 cm]
Here, $e \wedge f$ denotes the infimum of $e$ and $f$, which exists iff $e' \vee f'$ exists.
Two elements $e,f \in E$ are called \textit{orthogonal} if $e \leq f'$ or, equivalently, $f \leq e'$.
An element $e \neq 0$ in $E$ is called an \textit{atom} if there is no element $f$ in $E$ 
with $f \leq e$ and $0 \neq f \neq e$. 

The interpretation of this mathematical terminology is as follows: 
orthogonal events are exclusive, $e'$ is the negation of $e$, and
$e \vee f$ is the disjunction of the two exclusive events $e$ and $f$.

It is not assumed that $E$ is a \textit{lattice} (in a lattice, 
there is a smallest upper bound $e \vee f$ and largest lower bound $e \wedge f$
for any two elements $e$ and $f$).
If $E$ were a distributive lattice (i.e., $e \wedge (f \vee g) = (e \wedge f) \vee (e \wedge g)$ 
for all $e,f,g \in E$), it would become a Boolean lattice
or Boolean algebra. The orthomodular law is a weakening of the distributivity law. 

Classical probability theory uses Boolean lattices as mathematical structure 
for the random events, and it can be expected that those subsets of $E$,
which are Boolean lattices, behave classically. Therefore, a subset $E_0$ of $E$
is called \textit{compatible} if there is a Boolean lattice $B$ with $E_0 \subseteq B \subseteq E$.
Any subset with pairwise orthogonal element is compatible \cite{brabec1979compatibility}.
Two subsets $E_1$ and $E_2$ of $E$ are called \textit{compatible with each other}
if the union of any compatible subset of $E_1$ with any compatible subset of $E_2$
is a compatible subset of $E$. Note that this does not imply that $E_1$ or $E_2$ themselves 
are compatible subsets.

A subset of an orthomodular \underline{lattice} is compatible if each pair of elements in this subset 
forms a compatible subset. However, the pairwise compatibility of the elements
of a subset of an orthomodular partially ordered set 
does not any more imply the compatibility of this subset 
\cite{brabec1979compatibility}.

A quantum logical structure, which is more general than the orthomodular partially ordered sets,
has been used in Refs. \cite{niestegge2001non, niestegge2008approach}. 
This more general structure is sufficient when only compatible pairs of elements in the 
quantum logic are considered. However, compatible subsets with more than two elements 
will play an important role in this paper. 

A quantum logic is a purely algebraic structure, unfurling its full potential 
only when its state space has some nice properties which shall be 
considered in the next section.  
\\[0,5 cm]
\large
\textbf{3. Non-classical conditional probability}
\normalsize
\\[0,5 cm]
The states on the orthomodular partially ordered set $E$
are the analogue of the 
probability measures in classical probability theory, and 
conditional probabilities can be defined similar to 
their classical prototype. 
A \textit{state} $\rho$ allocates the probability $ \rho(f)$ 
with $0 \leq \rho(f) \leq 1$ to each event $f \in E$, 
is additive for orthogonal events, and $\rho(\mathbb{I})=1$.
It then follows that $\rho(f) \leq \rho(e)$ for any two events
$e,f \in E$ with $f \leq e$.

The \textit{conditional 
probability} of an event $f$ under another event $e$ is the 
updated probability for $f \in E$ after the outcome of
a first measurement has been the event $e \in E$; it is denoted 
by $ \rho(f | e) $. Mathematically, it is defined by the
conditions that the map $E \ni f \rightarrow \rho(f | e)$
is a state on $E$ and that it coincides with the classical
conditional probability for those $f$ which are compatible
with $e$. The second condition is equivalent to the  
identity $ \rho(f | e) = \rho(f)/\rho(e)$ for all events 
$f \in E$ with $f \leq e$. It must be assumed that $\rho(e) \neq 0$.

However, among the orthomodular partially ordered sets, 
there are many where no states or no conditional 
probabilities exist, or where the conditional probabilities 
are ambiguous. It shall now be assumed 
for the remaining part of this paper that 
\begin{enumerate}
\item[(F)] there is a state $\rho$ on $E$ with $\rho(e)\neq 0$ for each $e \in E$ with $e \neq 0$,
\item[(G)] $E$ possesses unique conditional probabilities, and
\item[(H)] the state space of $E$ is strong; i.e., if 
\newline
\hspace*{1,4 cm} $\left\{ \rho \ |\  \rho \mbox{ is a state with } \rho(f) = 1 \right\} $
\newline
\hspace*{1 cm} $\subseteq \left\{ \rho \ |\  \rho \mbox{ is a state with } \rho(e) = 1 \right\}$
\newline
holds for two events $e$ and $f$ in $E$, then $f \leq e$. 
\end{enumerate}

If $\rho$ is a state with $\rho(e) = 1$ for some event $e \in E$, 
then $ \rho(f | e) = \rho(f)$ for all $f \in E$. This follows from (G).

For some event pairs $e$ and $f$ in $E$, the conditional probability
does not depend on the underlying state; this means 
$\rho_1 (f|e) = \rho_2 (f|e)$ for all states $\rho_1$ and $\rho_2$
with $\rho_1 (e) \neq 0 \neq \rho_2 (e)$. This special conditional
probability is then denoted by $\mathbb{P} (f|e)$. The following 
two conditions are equivalent for an event pair $e,f \in E$:
\newline
\hspace*{0,2 cm}
(i) $\mathbb{P} (f|e)$ exists and $\mathbb{P} (f|e) = s$.
\newline
\hspace*{0,2 cm}
(ii) $\rho(e) = 1$ implies $\rho(f) = s$ for the states $\rho$ on $E$. 
\newline
Due to condition (H), $f \leq e$ holds for two events $e$ and $f$ in $E$
if and only if $ \mathbb{P} (e|f) = 1$. Moreover, $e$ and $f$ are orthogonal
if and only if $ \mathbb{P} (e|f) = 0$.
 
$\mathbb{P} (f|e)$ exists for all $f \in E$ if and only if $e$ is an \textit{atom}
(minimal event), which results in the atomic state
$\mathbb{P}_e$ defined by $\mathbb{P}_e (f) := \mathbb{P} (f|e)$.
This is the unique state allocating the probability value $1$ to the atom $e$.
For two atoms $e$ and $f$ in $E$, the following four identities are equivalent:
$\mathbb{P}_e(f) = 1$, $\mathbb{P}_f(e) = 1$, $\mathbb{P}_e = \mathbb{P}_f$, and $e = f$.
\\[0,5 cm]
\large
\textbf{4. Morphisms}
\normalsize
\\[0,5 cm]
In this section, the invariance of the special conditional 
probability $\mathbb{P}(\cdot|\cdot)$ under quantum logical morphisms is studied. 
In the proof of the main result, this will later replace the invariance of the
inner product under unitary transformations in the Hilbert space setting.

Suppose $E$ and $F$ are orthomodular partially ordered sets 
and $T:E \rightarrow F$ is an (algebraic) morphism 
(i.e., $Te_1 \leq Te_2$ for $e_1, e_2 \in E$ with $e_1 \leq e_2$, $T(e') = (Te)'$ for all $e \in E$ and $T \mathbb{I} = \mathbb{I}$). 
A dual transformation $T^{*}$,
mapping the states $\rho$ on $F$ to states $T^{*} \rho$  on $E$, is then defined by 
$(T^{*} \rho) (e) := \rho (T e)$ for $e \in E$.
In the case where both $E$ and $F$ possess unique conditional probabilities,
$$(T^{*}\rho)(e_2|e_1) = \rho(T e_2|T e_1)$$ 
holds for all events $e_1, e_2 \in E$ with $\rho(T e_1) \neq 0$.
To see this, consider the state $e \rightarrow \rho(T e|T e_1)$ on $E$;
the uniqueness of the conditional probability implies that it must
coincide with the state $e \rightarrow (T^{*}\rho)(e|e_1)$. 
\\[0,2 cm]
\textbf{Lemma 1:} Let $E$ and $F$ be orthomodular partially ordered sets, 
satisfying (F) and (G),
and let $T: E \rightarrow F$ be a morphism. 
\newline
(a) If $\mathbb{P} \left( e_2| e_1 \right)$
exists for two events $e_1$ and $e_2$ in $E$ with $T e_1 \neq 0$, then 
$\mathbb{P} \left( T e_2| T e_1 \right)$ exists and 
$$\mathbb{P} \left( T e_2| T e_1 \right) = \mathbb{P} \left( e_2| e_1 \right).$$
(b) If $T$ is an isomorphism, then 
$T^{*} \mathbb{P}_f = \mathbb{P}_{T^{-1}f}$ for the atoms $f$ in $F$.
\\[0,2 cm]
\textit{Proof}. (a) Suppose $\mathbb{P} \left( e_2| e_1 \right)$
exists for $e_1$ and $e_2$ in $E$. Then 
$\mathbb{P} \left( e_2| e_1 \right)$ = $(T^{*}\rho)(e_2|e_1)$ = $\rho(T e_2|T e_1)$
for all states $\rho$ on $F$ with $ (T^{*}\rho)(e_1) = \rho (T e_1) > 0$. Therefore,
$\mathbb{P} \left( T e_2| T e_1 \right)$ exists and is identical with
$\mathbb{P} \left( e_2| e_1 \right)$.
\newline
(b) Let $f$ be an atom in $F$. Then ${T^{-1}f}$ is an atom in $E$
and, by (a), $\mathbb{P}(e|T^{-1}f) = \mathbb{P}(Te|f) $ for $e \in E$.
Therefore, $\mathbb{P}_{T^{-1}f} = T^{*} \mathbb{P}_f$.
\\[0,5 cm]
\large
\textbf{5. The generalized no-cloning theorem}
\normalsize
\\[0,5 cm]
In this section, a quantum logic shall always be an orthomodular partially ordered set 
and shall satisfy (F), (G) and (H).
Suppose that $E$ is a quantum logic and that two copies of it are 
contained in the larger quantum logic $L$. This means that
there are two injective morphisms $\pi_1:E \rightarrow L$ and $\pi_2:E \rightarrow L$.
Moreover, suppose
\begin{enumerate}
\item[(I)] the subsets $\pi_1(E)$ and $\pi_2(E)$ of $L$ 
are compatible with each other and 
\item[(J)] $\pi_1(e) \wedge \pi_2(f)$ is an atom in $L$ for each pair of atoms $e$ and $f$ in $E$.
\end{enumerate}

The proof of the quantum mechanical no-cloning theorem rests upon 
the multiplicativity of the inner product for the Hilbert space tensor product. 
The following lemma provides the substitute for this in the more general setting.
\\[0,2 cm]
\textbf{Lemma 2:} Suppose $\mathbb{P}(e_2|e_1)$ and $\mathbb{P}(f_2|f_1)$
both exist for $e_1, e_2, f_1, f_2 \in E$. Then 
$$\mathbb{P}((\pi_1 e_2) \wedge (\pi_2 f_2)|(\pi_1 e_1) \wedge (\pi_2 f_1))$$
exists and
$$\mathbb{P}((\pi_1 e_2) \wedge (\pi_2 f_2)|(\pi_1 e_1) \wedge (\pi_2 f_1)) = \mathbb{P}(e_2|e_1) \mathbb{P}(f_2|f_1).$$
\textit{Proof}. Given the above assumptions, consider a state
$\rho$ on $L$ with 
$$\rho((\pi_1 e_1)\wedge (\pi_2 f_1)) = 1.$$
Then $\rho(\pi_1 e_1) = 1 = \rho(\pi_2 f_1)$.
Now define two states $\mu_1$ and $\mu_2$ on $E$ by
$$\mu_1(e) := \rho((\pi_1 e) \wedge (\pi_2 f_1)) 
\ \mbox{and}\ 
\mu_2(e) := \rho((\pi_1 e))$$
for $e \in E$.
Note that $\mu_1$ is a state due to (I).
Since $\mu_1(e_1) = 1 = \mu_2(e_1)$ and $\mathbb{P}(e_2|e_1)$ exists,
it follows that $\mathbb{P}(e_2|e_1) = \mu_1(e_2) = \mu_2(e_2)$ and 
$\mathbb{P}({e_2}'|e_1) = \mu_1({e_2}') = \mu_2({e_2}')$.
Thus 
$$\mathbb{P}(e_2|e_1) = \rho((\pi_1 e_2) \wedge (\pi_2 f_1)) = \rho((\pi_1 e_2))$$
and
$$\mathbb{P}({e_2}'|e_1) = \rho((\pi_1 {e_2}') \wedge (\pi_2 f_1)) = \rho((\pi_1 {e_2}')).$$
In the case $\mathbb{P}(e_2|e_1) > 0$, define the state $\nu$ on $E$ by
$$\nu(f) := \frac{\rho((\pi_1 e_2) \wedge (\pi_2 f))}{ \mathbb{P}(e_2|e_1) }$$
for $f \in E$.
Then $\nu(f_1) = 1$ and therefore 
$$\mathbb{P}(f_2|f_1) = \nu(f_2) = \frac{\rho((\pi_1 e_2) \wedge (\pi_2 f_2))}{\mathbb{P}(e_2|e_1)}.$$
In the case $\mathbb{P}(e_2|e_1) = 0$, it follows 
$\mathbb{P}({e_2}'|e_1) = 1$. Then $e_1 \leq {e_2}'$ and $e_2 \leq {e_1}'$.
Therefore, 
$$\rho((\pi_1 e_2) \wedge (\pi_2 f_2)) \leq \rho(\pi_1 e_2) \leq \rho(\pi_1 {e_1}') = 1 - \rho(\pi_1 e_1)= 0$$
and $\rho((\pi_1 e_2) \wedge (\pi_2 f_2)) = 0$.
In both cases, 
$$\rho((\pi_1 e_2) \wedge (\pi_2 f_2)) = \mathbb{P}(e_2|e_1) \mathbb{P}(f_2|f_1).$$
Since this holds for all states $\rho$ on $L$ with 
$\rho((\pi_1 e_1)\wedge (\pi_2 f_1)) = 1$,
it finally follows that
$$\mathbb{P}((\pi_1 e_2) \wedge (\pi_2 f_2)|(\pi_1 e_1) \wedge (\pi_2 f_1)) = \mathbb{P}(e_2|e_1) \mathbb{P}(f_2|f_1).$$
Note that the proof of Lemma 2 does neither require assumption (J) nor any tensor product construction, but instead only assumption(I). 
\\[0,2 cm]
A state $\rho $ on $L$ can be restricted to each one of the two copies of $E$ in $L$,  
resulting in the following two states on $E$:
$\pi_1^{*} \rho = \rho \pi_1$ and $\pi_2^{*} \rho = \rho \pi_2$. 
\\[0,2 cm]
\textbf{Lemma 3:} Let $e$ and $f$ be atoms in $E$ and $\rho $ a state on $L$. Then 
$\rho \pi_1 = \mathbb{P}_e$ 
and $\rho \pi_2 = \mathbb{P}_f$ if and only if 
$\rho = \mathbb{P}_{\left(\pi_1 e\right) \wedge \left(\pi_2 f \right)}$.
\\[0,2 cm]
\textit{Proof}. Assume $\rho \pi_1 = \mathbb{P}_e$ 
and $\rho \pi_2 = \mathbb{P}_f$. Then (I) implies
$$1 = \mathbb{P}_e e = \rho \pi_1 e = 
\rho \left(\left(\pi_1 e\right) \wedge \left(\pi_2 f \right)\right) + 
\rho \left(\left(\pi_1 e\right) \wedge \left(\pi_2 f' \right)\right)$$
and 
$$0 \leq \rho \left(\left(\pi_1 e\right) \wedge \left(\pi_2 f' \right)\right) \leq \rho \pi_2 f' 
= \mathbb{P}_f f' = 0.$$ 
Therefore 
$1 = \rho \left(\left(\pi_1 e\right) \wedge \left(\pi_2 f \right)\right)$
and, since $\left(\pi_1 e\right) \wedge \left(\pi_2 f \right)$ is an atom in $L$, 
$$\rho = \mathbb{P}_{\left(\pi_1 e\right) \wedge \left(\pi_2 f \right)}.$$
Now assume $\rho = \mathbb{P}_{\left(\pi_1 e\right) \wedge \left(\pi_2 f \right)}$
and $a \in E$. Then by Lemmas 1(a) and 2
\\[0,2 cm]
\hspace*{1,2 cm} $\rho \pi_1 a = \mathbb{P} ((\pi_1 a) | (\pi_1 e) \wedge (\pi_2 f))$
\\[0,2 cm]
\hspace*{2 cm} $= \mathbb{P} ((\pi_1 a) \wedge (\pi_2 \mathbb{I}) | (\pi_1 e) \wedge (\pi_2 f))$
\\[0,2 cm]
\hspace*{2 cm} $= \mathbb{P} (\pi_1 a| \pi_1 e) \mathbb{P} (\pi_2 \mathbb{I}| \pi_2 f)$
\\[0,2 cm]
\hspace*{2 cm} $= \mathbb{P} (\pi_1 a| \pi_1 e)$
\\[0,2 cm]
\hspace*{2 cm} $= \mathbb{P} (a|e) = \mathbb{P}_e a$
\\[0,2 cm]
The second identity $\rho \pi_2 = \mathbb{P}_f$ follows in the same way.
\\[0,2 cm]
Now suppose that $C$ is a set of atoms in $E$ and that $f$ is a fixed atom in $E$,
that the local state on the first copy of $E$ is any element in $\left\{\mathbb{P}_e | e \in C\right\}$
and that the local state on the second copy of $E$ is $\mathbb{P}_f$.
For the state $\rho$ on $L$ this means that
$\rho \pi_1 = \mathbb{P}_e$ for some unknown $e \in C$
and $\rho \pi_2 = \mathbb{P}_f$.

In the usual quantum mechanical setting, the cloning is performed 
by a unitary transformation on the Hilbert space tensor product. In this paper, it
shall be performed by an automorphism of $L$; cloning means that the automorphism 
transforms the initial local state on the second copy of $E$ to a copy of the unchanged local state 
on the first copy of $E$. After the transformation, both copies of $E$ are in the same local state 
and this is the local state on the first copy before the transformation. 
\\[0,2 cm]
\textbf{Definition 1:} A cloning transformation for $\left\{\mathbb{P}_e | e \in C\right\}$ 
is an automorphism $T$ of the quantum logic $L$ such that
$(T^{*} \rho) \pi_1 = \rho \pi_1 = (T^{*} \rho) \pi_2$ holds
for the states $\rho $ on $L$ with $\rho \pi_1 \in \left\{\mathbb{P}_e | e \in C\right\}$ 
and $\rho \pi_2 = \mathbb{P}_f$.
\\[0,2 cm]
\textbf{Theorem 1:} A cloning transformation $T$ for $\left\{\mathbb{P}_e | e \in C\right\}$ 
exists only if the atoms in $C$ are pairwise orthogonal.
\\[0,2 cm]
\textit{Proof}. Assume $T$ is a cloning transformation for 
$\left\{\mathbb{P}_e | e \in C\right\}$. Note that,
by Lemma 3, $\rho \pi_1 = \mathbb{P}_e$ 
and $\rho \pi_2 = \mathbb{P}_f$ holds for the states 
$\rho = \mathbb{P}_{\left(\pi_1 e\right) \wedge \left(\pi_2 f \right)}$ 
with $e \in C$ and, furthermore, 
$(T^{*} \rho) \pi_1 = \mathbb{P}_e = (T^{*} \rho) \pi_2$ with $e \in C$
implies $T^{*} \rho = \mathbb{P}_{\left(\pi_1 e\right) \wedge \left(\pi_2 e \right)}$.
Therefore by Lemma 1(b)
$$ \mathbb{P}_{\left(\pi_1 e\right) \wedge \left(\pi_2 e \right)} = T^{*} \mathbb{P}_{\left(\pi_1 e\right) \wedge \left(\pi_2 f \right)}
= \mathbb{P}_{T^{-1}\left(\left(\pi_1 e\right) \wedge \left(\pi_2 f \right)\right)}$$
and thus 
$$ T^{-1}\left((\pi_1 e) \wedge (\pi_2 f) \right)= \left((\pi_1 e) \wedge (\pi_2 e)\right)$$
for each $e \in C$.
Now assume $e_1, e_2 \in C$ and consider 
$$ \mathbb{P} \left( (\pi_1 e_2) \wedge (\pi_2 f)| (\pi_1 e_1) \wedge (\pi_2 f) \right).$$
On the one hand, the repeated application of Lemmas 1(a) and 2 yields 
\\[0,2 cm] \hspace*{0,5 cm}
$ \mathbb{P} \left( (\pi_1 e_2) \wedge (\pi_2 f)| (\pi_1 e_1) \wedge (\pi_2 f) \right) $
\\[0,2 cm] \hspace*{1 cm}
$ = \mathbb{P} \left( T^{-1}\left((\pi_1 e_2) \wedge (\pi_2 f)\right)| T^{-1}\left((\pi_1 e_1) \wedge (\pi_2 f) \right)\right) $
\\[0,2 cm] \hspace*{1 cm}
$ = \mathbb{P} \left( ( \pi_1 e_2) \wedge (\pi_2 e_2) | (\pi_1 e_1) \wedge (\pi_2 e_1) \right) $
\\[0,2 cm] \hspace*{1 cm}
$ = \mathbb{P} \left( \pi_1 e_2 | \pi_1 e_1 \right) \mathbb{P} \left( \pi_2 e_2 | \pi_2 e_1 \right) $
\\[0,2 cm] \hspace*{1 cm}
$ = \mathbb{P} \left( e_2 | e_1 \right) \mathbb{P} \left( e_2 | e_1 \right) $
\\[0,2 cm] \hspace*{1 cm}
$ = \left(\mathbb{P} \left( e_2 | e_1 \right)\right)^{2} $
\\[0,2 cm]
and, on the other hand, 
\\[0,2 cm] \hspace*{0,5 cm}
$ \mathbb{P} \left( (\pi_1 e_2) \wedge (\pi_2 f)| (\pi_1 e_1) \wedge (\pi_2 f) \right)$
\\[0,2 cm] \hspace*{1 cm}
$ = \mathbb{P} \left( \pi_1 e_2 | \pi_1 e_1 \right) \mathbb{P} \left(\pi_2 f| \pi_2 f \right)$
\\[0,2 cm] \hspace*{1 cm}
$ = \mathbb{P} \left( e_2| e_1 \right) \mathbb{P} \left( f|f) \right) $
\\[0,2 cm] \hspace*{1 cm}
$ = \mathbb{P} \left( e_2| e_1 \right).$
\\[0,2 cm]
Therefore, $ \left(\mathbb{P} \left( e_2 | e_1 \right)\right)^{2} = \mathbb{P} \left( e_2| e_1 \right)$ 
and $\mathbb{P} \left( e_2| e_1 \right) \in \left\{0,1\right\}$. This means that $e_1$ and $e_2$ 
are either orthogonal or identical.
\\[0,2 cm]
If $L$ is a finite Boolean algebra (i.e., classical), different atoms
are orthogonal and a cloning transformation $T$ is defined by 
extending to $L$ the following permutation of the atoms in $L$:
$T((\pi_1 e) \wedge (\pi_2 e)) = (\pi_1 e) \wedge (\pi_2 f)$ and 
$T((\pi_1 e) \wedge (\pi_2 f)) = (\pi_1 e) \wedge (\pi_2 e)$ for all $e \in C$,
$Td = d$ for the other atoms $d$ in $L$. However, 
non-orthogonal atoms are quite characteristic of quantum mechanics
and Theorem 1 rules out that the corresponding atomic states can be cloned.
\\[0,5 cm]
\large
\textbf{6. Quantum mechanics}
\normalsize
\\[0,5 cm]
Quantum mechanics uses a special quantum logic; 
it consists of the self-adjoint projection operators on a 
Hilbert space $H$ and is an orthomodular lattice. Compatibility 
here means that the self-adjoint projection operators
commute. Conditions (F) and (H) in section 3 are satisfied, and 
the unique conditional probabilities exist (G) 
unless the dimension of $H$ is two \cite{niestegge2001non}.
Moreover,  it has been shown in Ref. \cite{niestegge2001non} that,
with two self-adjoint projection operators $e$ and $f$ on $H$, the
conditional probability has the shape
$$\rho(f|e) = \frac{trace(aefe)}{trace(ae)} = \frac{trace(eaef)}{trace(ae)}$$
for a state $\rho$ defined by the statistical operator $a$ (i.e., $a$
is a self-adjoint operator on $H$ with non-negative spectrum and $trace(a)=1$). 
The above identity reveals that
conditionalization becomes identical with the state transition 
of the L\"uders - von Neumann measurement process. Therefore, 
the conditional probabilities defined in section 3 can be 
regarded as a generalized mathematical model of projective quantum measurement.

$\mathbb{P}(f|e)$ exists with $\mathbb{P}(f|e) = s$ if and only if the operators
$e$ and $f$ satisfy the algebraic identity $efe = se$.
This transition probability between the outcomes of two consecutive measurements
is independent of any underlying state and results from the algebraic structure
of the quantum logic.

The atoms are the self-adjoint projections on the one-dimensional subspaces 
of $H$; if $e$ is an atom and $\xi$ a normalized vector in the 
corresponding one-dimensional subspace, then 
$\mathbb{P}(f|e) = \left\langle \xi| f \xi\right\rangle$.
The atomic states thus coincide with the quantum mechanical pure states or vector states.
Their general non-orthogonality is quite characteristic of quantum mechanics.  

The quantum mechanical model of a composite system consisting of two copies 
is the Hilbert space tensor product $H \otimes H$. 
The self-adjoint projection operators $e$ on $H$ are mapped 
to two copies on $H \otimes H$ by
$\pi_1(e):=e \otimes \mathbb{I}$ and $\pi_2(e):= \mathbb{I} \otimes e$.
Note that (I) and (J) are then satisfied.
Time evolutions of the composite system are described by 
unitary transformations of $H \otimes H$. Therefore, 
the cloning operation should be such a transformation. It defines
an automorphism of the quantum logic of $H \otimes H$.

Theorem 1 thus includes the quantum mechanical no-cloning theorem 
for pure states as a special case. Instead of 
the Hilbert space and tensor product formalism, 
Theorem 1 requires only a few very basic principles; these are
the existence and the uniqueness of the conditional probabilities 
and the existence of two compatible copies of the system in a larger system.
Nevertheless, the proof 
of the no-cloning theorem in the quantum mechanical Hilbert space formalism
can be mimicked, 
replacing the Hilbert space inner product $\left\langle \  | \ \right\rangle$ by the specific state-independent 
conditional probability $\mathbb{P}(\ |\ )$.

Theorem 1 considers only the atomic or pure states, while other approaches 
to a generalized no-cloning or no broadcasting theorem 
\cite{barnum2006cloning, barnum2007generalized, :/content/aip/journal/jmp/50/10/10.1063/1.3245811}
include the mixed states. On the other hand, these approaches 
are restricted to finite-dimensional theories or universal cloning
and need an explicit tensor product construction.

\bibliographystyle{abbrv}
\bibliography{Literatur}

\end{document}